\documentclass[prb,twocolumn,amsmath,amssymb,superscriptaddress]{revtex4}

\usepackage[dvipdfmx]{graphicx}
\usepackage{pdfpages}
\usepackage{dcolumn}
\usepackage{bm}
\usepackage{color}
\usepackage{amssymb}

\begin{document}

\title{
Ground-state phase diagram of anisotropically interacting Heisenberg-$\Gamma$ models on a honeycomb lattice
}

\author{Takafumi Suzuki}
\author{Takuto Yamada}
\author{ Sei-ichiro Suga}

\affiliation{Graduate School of Engineering, University of Hyogo, Himeji 671-2280, Japan}

\date{\today}

\begin{abstract} 
In this paper, we investigate the ground-state phase diagram of the $S=1/2$ Heisenberg-$\Gamma$ model on a honeycomb lattice by dimer series expansion and exact diagonalization.
We focus on the effects of the anisotropy of the interactions; by tuning the coupling constants, the system changes between the isolated dimer and the spin-chain models. 
We find that, in the spin-chain limit, there are three kinds of states: a Tomonaga-Luttinger liquid and two magnetically long-range-ordered states. 
All three states become two-dimensional long-range ordered states by the infinitesimal interchain interaction except for the case where the Heisenberg interaction is much weaker than the off-diagonal symmetric ($\Gamma$) interaction.
Starting from the isolated dimer limit, a triplet dimer phase survives up to the isotropically interacting system in a large part of the phase diagram where the Heisenberg and $\Gamma$ interactions are ferromagnetic and antiferromagnetic, respectively. 
Otherwise, a phase transition to a magnetically ordered phase occurs before the interaction becomes isotropic.
This indicates that the quantum spin liquid proposed in the $\Gamma$ model [A. Catuneanu {\it et al}., npj Quantum Mater. {\bf 3}, 23 (2018)] is unstable against the anisotropy of the interactions.
\end{abstract}

\preprint{APS/123-QED}

\maketitle

\section{\label{sec:level1}Introduction}
Recently, much condensed matter physics researches have been focused on realizing quantum spin liquids.
Kitaev spin liquid\cite{Kitaev} is one of the most studied, since several magnets including ${\rm Na_2IrO_3}$ and $\alpha$-${\rm RuCl_3}$ emerged as promising candidates for the Kitaev model \cite{Khaliullin,SKChoi,FYe,JChaloupka,RComin,Takagi,Plump,Kubota,Sandilands,Do,DHirobe,Sears,Banerjee1,Banerjee2,Banerjee3}.
So far, effective magnetic models have been proposed by employing various {\it ab-initio} and {\it ab-initio}-guided calculations \cite{RComin,Katukuri2014,Yamaji2014,Sizyuku2014,Winter2016,WinterRev,Plump,LJanssenPRB,Majumder2015,Sandilands2016,HSKim2016,WWangPRB,SuzukiPRB2018,PLaurell2020}; they indicate that 
 these materials possess  the Heisenberg ($J$) and symmetric off-diagonal ($\Gamma$) couplings in addition to Kitaev ($K$) interactions.
Moreover, in real materials, these interactions include a spatial anisotropy \cite{Yamaji2014,Sizyuku2014,HSKim2016,Yamaji2018-1,Yamaji2018-2}, which plays a significant role in understanding Kitaev physics.
The simplest example may be the effect of the spatially anisotropic interaction in the pure Kitaev model. 
In the pure Kitaev model on a honeycomb lattice, the gapless Kitaev spin liquid undergoes a phase transition to gapped spin liquid related with the ground state of the toric code model\cite{Kitaev}.
In the Kitaev-$\Gamma$ model considering both Kitaev and $\Gamma$ interactions on the honeycomb lattice, when the spatial anisotropy of the interactions is strong, the Kitaev spin liquid connects adiabatically to the spin liquid appearing in $|\Gamma/K|\gg1$ \cite{Yamaji2018-1,Yamaji2018-2}.
The presence of this spin liquid phase is under debate, because variational Monte Carlo study predicts that one proximate Kitaev spin liquid phase appears in  $0.2 \lessapprox \Gamma/|K| \lessapprox 0.6$ and the zigzag phase is stable in $0.6 \lessapprox \Gamma/|K|$ \cite{WangPRL2019}.
These findings have prompted us to investigate the ground-state properties of relevant models for Kitaev magnets that include $K$, $\Gamma$, and $J$ interactions.

In a previous study on the Kitaev-$\Gamma$ model\cite{TYamada}, we clarified that a dimerized state survives up to the isotropically interacting system under fixed $\Gamma/K$.
From the resulting phase diagram, we inferred that the isotropically interacting system is located at the phase boundary in the Kitaev-$\Gamma$ model with $\Gamma>0$. 
We attributed this to a frustration effect between the Kitaev and the $\Gamma$ interactions.
In contrast, nearest-neighbor Heisenberg interactions on the honeycomb lattice themselves are free of frustration, favoring the magnetically ordered state.
Thus, we expect the ground-state phase diagram of the Heisenberg-$\Gamma$ model to possess more different features than that of the Kitaev-$\Gamma$ model.
Although the Heisenberg interaction is weak in $\alpha$-RuCl$_3$, the ground-state phase diagram of the Heisenberg-$\Gamma$ model provides insights on the possibility of the quantum spin liquid argued in the Kitaev-$\Gamma$ model with the strong $\Gamma$ interaction \cite{Yamaji2018-1}.

In this paper, we investigate the ground-state phase diagram of the $S=1/2$ Heisenberg-$\Gamma$ model on the honeycomb lattice by varying the anisotropy of the interactions. The Hamiltonian is described by
\begin{eqnarray}
&{\mathcal H}&=t ({\mathcal H}_{X} + {\mathcal H}_{Y})+(1-2t){\mathcal H}_Z,
\label{Ham1}
\end{eqnarray}
where
\begin{eqnarray}
&{\mathcal H}_{X,Y}&=\sum_{{\langle ij \rangle}_{\gamma=X,Y}}  \left[ J {\boldsymbol S_i}\cdot {\boldsymbol S_j} + \Gamma \left( {S_i}^{\alpha} {S_j}^{\beta} + {S_i}^{\beta} {S_j}^{\alpha} \right)  \right]
\end{eqnarray}
and
\begin{eqnarray}
&{\mathcal H}_{Z}&=\sum_{{\langle ij \rangle}_{Z}}  \large[ J {\boldsymbol S_i}\cdot {\boldsymbol S_j} + \Gamma \left( {S_i}^{x} {S_j}^{y} + {S_i}^{y} {S_j}^{x} \right)  \large].
\end{eqnarray}
Here, $\langle ij \rangle_{\gamma = {X,Y,Z}}$ represents a nearest-neighbor pair on the $\gamma$ bond of the honeycomb lattice, 
and both $\alpha$ and $\beta$ are the different spin components from the $\gamma$ component. 
The anisotropy of the interactions is introduced by $t$. 
%

This model includes the two limits: the isolated dimer model ($t=0$) and the spin chain model ($t=1/2$).
Starting from the dimerized limit at $t=0$, we examine the ground state up to the isotropically interacting system ($t=1/3$) by employing series expansions; these are based on graph theories and can include systematically higher-order terms \cite{textbook,JOitmaa}. 
We adopt dimer series expansions, where the interactions on specific bonds are included in the initial state.
Starting from isolated dimers on the $Z$ bond [Fig. \ref{Phasediagram}(a)], we consider the interactions between the dimers are incorporated perturbatively.
We calculate the ground-state energy and evaluate its first and second derivatives numerically; furthermore, we also calculate these quantities, employing  the numerical exact diagonalization method (ED) for a 24-site cluster as shown in Fig. \ref{Phasediagram}(b).  
Thus, series expansions and ED can be used complementarily.
To understand the ground-state properties in the chain limit at $t=1/2$, we employ the ED and the density-matrix-renormalization-group (DMRG) calculations.
Starting from the chain limit as shown in Fig. \ref{Phasediagram}(c), we also calculate the ground-state energy and its first and second derivatives with the 24-site ED by increasing the interchain interactions.
Finally, combining all the results, we obtain the ground-state phase diagram of the anisotropically interacting Heisenberg-$\Gamma$ model between the spin chain limit and the isolated dimer limit.

The rest of the paper is organized as follows.
We show the ground-state phase diagram and the features of each phases in Sec. \ref{sec:level2}. 
We present the details on inspection of the phase diagram in Sec. \ref{sec:level4}.
First, we discuss the ground state at the three characteristic points where the model becomes equivalent to the spin-chain, the isolated-dimer, and the isotropically interacting models.
Next, we explore the stable states when the interchain/interdimer interactions are strengthened with respect to the Heisenberg and $\Gamma$ interactions. 
We find that, when $t$ is increased, depending on the ratio of Heisenberg and $\Gamma$ interactions, the dimer state either undergoes a phase transition at an anisotropic interaction, or survives up to the isotropically interacting system where a phase transition occurs. 
Based on the phase diagram, we argue that the quantum spin liquid\cite{Yamaji2018-1,Yamaji2018-2} anticipated near the $\Gamma$ model is absent.  
Finally, we summarize this paper in Sec. \ref{sec:level5}.

\section{\label{sec:level2}Overview of the Phase Diagram}
\subsection{Outline of dimer series expansions}

When $t=0$, the system is described by the isolated dimer model.
For $0 \le t \le 1/3$, we investigate the ground-state properties by dimer series expansions \cite{textbook}, treating ${\mathcal H}_Z$ as the unperturbed term and ${\mathcal H}_{X}+{\mathcal H}_Y$ as the perturbation.
The schematic of dimer series expansions is shown in Fig. \ref{Phasediagram}(a).
We rewrite the original Hamiltonian (\ref{Ham1}) as
\begin{eqnarray}
{\mathcal H'}={\mathcal H}_Z+ \lambda ({\mathcal H}_{X} + {\mathcal H}_{Y}),
 \label{Ham2}
\end{eqnarray}
where $\lambda=t/(1-2t)$.
Starting from the isolated initial dimers on the $Z$ bonds, we perform dimer series expansions with respect to $\lambda ({\mathcal H}_{X}+{\mathcal H}_{Y})$ up to the eighth order. 
For the unperturbed Hamiltonian ${\mathcal H}_Z$, the lowest energy state on each bond is selected from one of the following four candidates: singlet dimer $|s\rangle=(|\uparrow\downarrow\rangle - |\downarrow\uparrow\rangle)/\sqrt{2}$,  triplet dimers $|t_0\rangle=(|\uparrow\downarrow\rangle + |\downarrow\uparrow\rangle)/\sqrt{2}$, $|t_x\rangle=(|\uparrow\uparrow\rangle - i|\downarrow\downarrow\rangle)/\sqrt{2}$,  and  $|t_y\rangle=(|\uparrow\uparrow\rangle + i|\downarrow\downarrow\rangle)/\sqrt{2}$, where the up and down arrows indicate the spin-up and spin-down states, respectively.
We perform the dimer series expansions by adopting the optimal initial dimer for $\theta$.
We calculate the ground-state energy per unit cell, $E$, as well as its first and second derivatives, $\partial E/\partial t$ and $\partial^2 E/\partial t^2$. We also calculate these quantities by the 24-site ED.  
Starting from $t=0\ (\lambda=0)$, we discuss the ground-state properties toward the isotropically interacting system at $t=1/3\ (\lambda=1)$.



\subsection{Phase diagram}
\begin{figure*}[htb]
\begin{center}
\includegraphics[width=0.95\hsize]{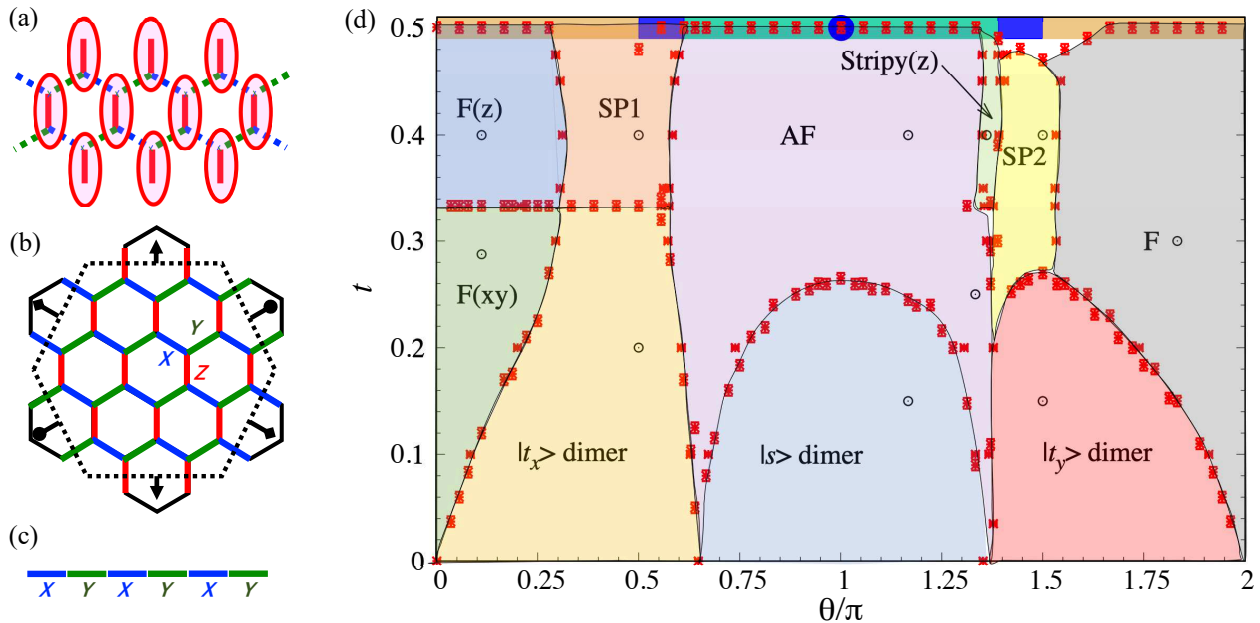}
\hspace{0pc}
\vspace{0pc}
\caption{(Color online) \label{Phasediagram} 
(a) Schematic of the dimer series expansions described in Sec. \ref{sec:level2}. 
Ellipsoids designate the initial spin dimers on the $Z$ bonds. 
Green and blue dotted lines denote the same strength interactions on the $X$ and $Y$ bonds.
(b) The $24$-site cluster. Periodic boundary conditions are applied along the dotted lines with common symbols. 
The $X, Y$, and $Z$ bonds are denoted by green, blue, and red colors, respectively. 
(c) Schematic picture of the spin chain model.
(d) Ground-state phase diagram determined from the 24-site ED for the Heisenberg-$\Gamma$ model, where $J=-\cos \theta$ and $\Gamma=\sin \theta, \;(0 \leq \theta < 2\pi)$.
The values $t=0,1/3$, and $1/2$ describe the isolated dimer, isotropically interacting, and spin chain models, respectively. 
All red asterisks mean that $\partial_{t,\theta} E$ shows a jump and/or ${\partial_{t,\theta}}^2 E$ shows a dip. 
Error bars are smaller than the symbol sizes. Thin solid lines and curves are guide to eyes.
"F (AF)" represent the ferromagnetic (antiferromagnetic) long-range ordered phase. 
In SP1 and SP2, weak antiferromagnetic and ferromagnetic correlations exist on the $Z$ bonds, respectively.
Characters in parentheses correspond to the spin component that shows the long-range order. 
Open symbols correspond to static structure factors shown in Fig. \ref{SQ}.
At $t=1/2$, lines colored by orange, green, and blue correspond to the ferromagnetic long-range ordered, antiferromagnetic long-range ordered, and Tomonaga-Luttinger (TL) liquid phases, respectively. 
Because the phase boundaries are determined by the 24-site ED, the noncolored area around $t=0.5$ and $\theta/\pi=3/2$ is expected to disappear in the thermodynamic limit.
It is also difficult to clarify the details of the other noncolored area in $0.2<t<1/3$ around $\theta/\pi=1.35$ because of the small system size.}
\end{center}
\end{figure*}
\begin{figure*}[htb]
\begin{center}
\includegraphics[width=0.9\hsize]{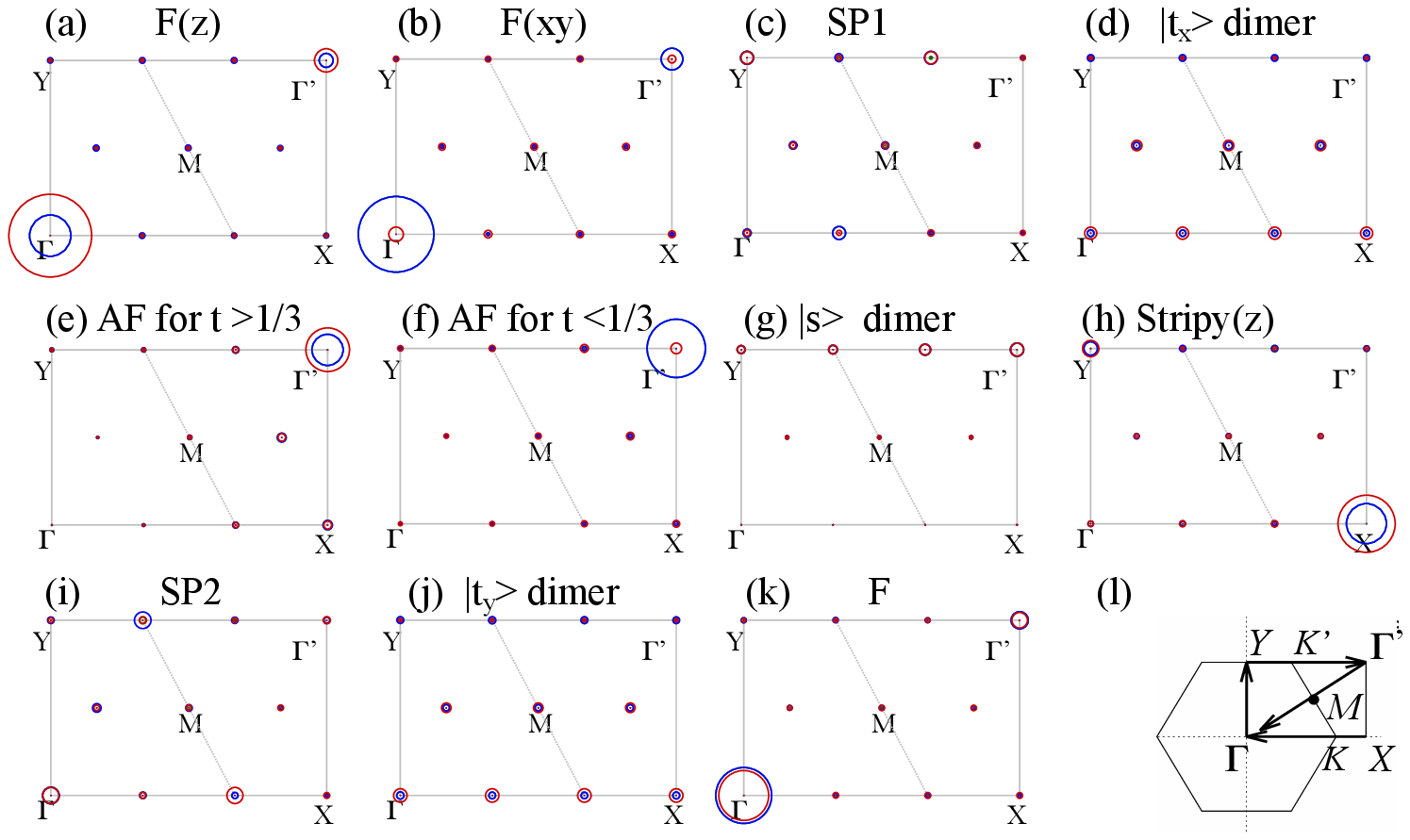}
\hspace{0pc}
\vspace{0pc}
\caption{(Color online) \label{SQ} Static structure factors $S({\boldsymbol Q})$ in each phase shown in Fig.  \ref{Phasediagram}(d): (a) F(z), (b) F(xy), (c) SP1, (d) $|t_x\rangle$ dimer, (e) AF for $t>1/3$, (f) AF for $t<1/3$, (g) $|s\rangle$ dimer, (h) Stripy(z), (i) SP2, (j) $|t_y\rangle$ dimer, and (k) F. (l) Brillouin zone of the four-sublattice system on honeycomb lattice. Green, blue, and red circles correspond to x, y, and z components of $S({\boldsymbol Q})$, respectively; the circle areas are proportional to the intensity. The results for the $x$ component are covered by those of the $y$ component.}
\end{center}
\end{figure*}

In Fig. \ref{Phasediagram}(d), we summarize the ground-state phase diagram of the Heisenberg-$\Gamma$ model.
All of phase boundaries are settled by the first and second derivatives of the ground-state energy of the 24-site ED.
In this paper, we parametrize the Heisenberg and $\Gamma$ interactions with $\theta:J=-\cos \theta$ and $\Gamma=\sin \theta \;(0 \leq \theta < 2\pi)$. 
Thus, $\theta=0$ ($\pi$) describes the honeycomb-lattice Heisenberg model constructed by ferromagnetic (antiferromagnetic) interactions, while $\theta=\pi/2 \; (3\pi/2)$ describes the $\Gamma$ model where only the antiferromagnetic (ferromagnetic) $\Gamma$ interaction between nearest-neighbor spins is present. 
Note that the antiferromagnetic (ferromagnetic) $\Gamma$ interaction means positive (negative) $\Gamma$.

The obtained phase diagram indicates the existence of ten phases: three ferromagnetically ordered [F(z), F(xy), and F], three dimerized ($|t_x\rangle$ dimer,  $|s\rangle$ dimer, and  $|t_y\rangle$ dimer), two spiral (SP1 and SP2), one antiferromagnetically ordered (AF), and one stripy [Stripy(z)] phases.
In Fig. \ref{SQ}, we show the static structure factors $S({\boldsymbol Q})$, corresponding to each of these ordered phases.

For $0<\theta/\pi<0.3$, we find the two ferromagnetically ordered phases for which the ordering spin component depends on $t$.  
For $0<t<1/3$, the $x$ (and $y$) components of $S({\boldsymbol Q})$ display the largest intensity at the $\Gamma$ point [Fig. \ref{SQ}(a)]; this indicates that the spins are ferromagnetically ordered on the $(x, y)$ plane in the spin space. 
On the other hand, for $1/3<t<1/2$, the $z$ component of $S({\boldsymbol Q})$ exhibits the largest intensity also at the $\Gamma$ point, as shown in Fig. \ref{SQ}(b), revealing that  the spins are ferromagnetically ordered parallel to the $z$-axis in the spin space. 
F(z) and F(xy) are separated by the first-order phase transition, where $E$ shows the cusp illustrated in Fig. \ref{fig:EDChain_to_iso} (a).

For $0.317<\theta/\pi<0.572$, the $|t_x\rangle$-dimer phase undergoes a phase transition to SP1 at $t=1/3$.  
In SP1, the intensity of $S({\boldsymbol Q})$ [Fig. \ref{SQ}(c)] at each wave vector has the same order, although it exhibits the relatively large value at the $Y$ point.
This implies the development of the incommensurate spin-spin correlation.
Since the present result is based on the 24-site ED, it is difficult to analyze the system size dependence and discuss the properties in the thermodynamic limit.
Thus, the nature of SP1 in the thermodynamic limit leaves our future study.
For $t\le1/3$, the $|t_x\rangle$ dimer is stable, and no prominent peak appears in $S({\boldsymbol Q})$ [Fig. \ref{SQ}(d)]. 
Since as will be discussed in Sec. III D no clear signs of phase transitions are observed for $0.317<\theta/\pi<0.572$ and $0\le t < 1/3$, the $|t_x\rangle$ dimer starting from $t=0$ survives up to $t=1/3$.
This means that the $\Gamma$ spin liquid\cite{Yamaji2018-1,Yamaji2018-2} is unstable against both the anisotropy of the interactions and the Heisenberg interaction.
This result is similar to our previous result\cite{TYamada} for the Kitaev-$\Gamma$ model, where the $|t_x\rangle$-dimer phase survived up to $t=1/3$ for $0.184 < \theta'/\pi < 0.583$ ($K=-\cos \theta'$ and $\Gamma=\sin \theta'$).
In the Kitaev-$\Gamma$ model, the spin liquid near the antiferromagnetic $\Gamma$ model is unstable against the anisotropy of the interaction.

AF appears for $0.6 \lessapprox \theta/\pi \lessapprox 1.35$ at $t \approx1/3$ and the $|s \rangle$-dimer phase appears in $t \lessapprox 0.25$. 
In AF, the $z$ component of $S({\boldsymbol Q})$ exhibits the largest intensity at the $\Gamma'$ point for $1/3\lessapprox t<1/2$; the $x$ and $y$ components exhibit the largest intensity at the $\Gamma'$ point for $t \lessapprox 1/3$ [Figs. \ref{SQ}(e) and \ref{SQ}(f)]. 
No other phase transition appears between these two AFs.
Note that since SU(2) symmetry is satisfied at $\theta/\pi=1$ irrespective of $t$, the intensity profiles of these three components become identical.
In the $|s \rangle$-dimer phase, no prominent peaks appear as shown in Fig. \ref{SQ}(g).

For $t>1/3$ and at $\theta/\pi \approx 1.35$, one stripy\cite{Khaliullin} phase appears, namely, Stripy(z).
In Stripy(z), the largest intensity of $S({\boldsymbol Q})$ appears at the $X$ point in the $z$ component [Fig. \ref{SQ}(h)], revealing that the $S^z$ component exhibits the stripy order.
Stripy(z) extends in the lengthwise thin region.
Since the present result is based on the 24-site ED calculations, we cannot neglect the finite-size effect arising upon the emergence of this phase, and further calculations are needed for a definite conclusion.
At $\theta/\pi \approx 1.45$, the other spiral phase (SP2) appears for $0.25 < t < 0.48$.
In SP2, the relatively largest intensity in $S({\boldsymbol Q})$ appears at the $\Gamma$ point [Fig. \ref{SQ}(i)]. 
However, its value is almost comparable to those at the $K$ and $K'$ points.
To clarify the details of SP2 in the thermodynamic limit, further calculations beyond the 24-site ED are also needed.

As shown in Fig. \ref{Phasediagram}(d), the $|t_y\rangle$-dimer phase shows the phase transitions to the ferromagnetic long-range-ordered phase at $t_c<1/3$.
Typical results for $S({\boldsymbol Q})$ in the $|t_y\rangle$-dimer phase and the ferromagnetic long-range-ordered phase are shown in Figs. \ref{SQ}(j) and \ref{SQ}(k), respectively.
The absence of the $|t_y \rangle$-dimer phase at $t=1/3$ is explained similarly to that in the Kitaev-$\Gamma$ model \cite{TYamada}: the sign combination between the Heisenberg and $\Gamma$ interactions. 
When the Heisenberg and the $\Gamma$ interactions have the same signs, the frustration between these interactions is free.
Thus, simple magnetic ordering is favored, such as the ferromagnetic order and the N\'eel order.
Actually, for $\theta/\pi>3/2$, the Heisenberg and $\Gamma$ interactions possess the same signs, which results in the ferromagnetic order. 
When the signs of the $\Gamma$ and Heisenberg interactions are opposite, they cause frustration.  
In fact, for $0 \le \theta/\pi \lessapprox 1/2$ and $1 \lessapprox \theta/\pi \lessapprox 3/2$, the simple magnetic order tends to be suppressed and the various other phases appear.

\section{\label{sec:level4} Details of Inspection of the Phase Diagram}
First, we explain the results obtained at three characteristic points, namely $t=0$, $1/3$, and $1/2$ where the systems are described by the isolated dimer, the isotropically interacting, and the spin chain models, respectively.
We then explain the results for $0<t<1/3$ and $1/3<t<1/2$.

\subsection{Isolated dimer model: $t=0$}
When the system is described by isolated dimers on the $Z$ bonds, the ground state is described by the direct product  of the dimers selected from the following four candidates: the singlet dimer $|s\rangle$, and the triplet dimers $|t_0\rangle$, $|t_x\rangle$,  and  $|t_y\rangle$, depending on $\theta$. 
By increasing $\theta/\pi$ from zero to two, the ground state undergoes phase transitions from $|t_x\rangle$ to $|s\rangle$ at $\theta=\pi-\arctan 2$ and from $|s\rangle$ to $|t_y\rangle$ at $\theta=\pi+\arctan 2$.
Notice that, in the Heisenberg-$\Gamma$ isolated-dimer model, the $|t_0\rangle$-dimer state is always eliminated from the possible ground-state selection.

\subsection{Spin chain model: $t=1/2$}
\begin{figure}[htb]
\begin{center}
\includegraphics[width=0.9\hsize]{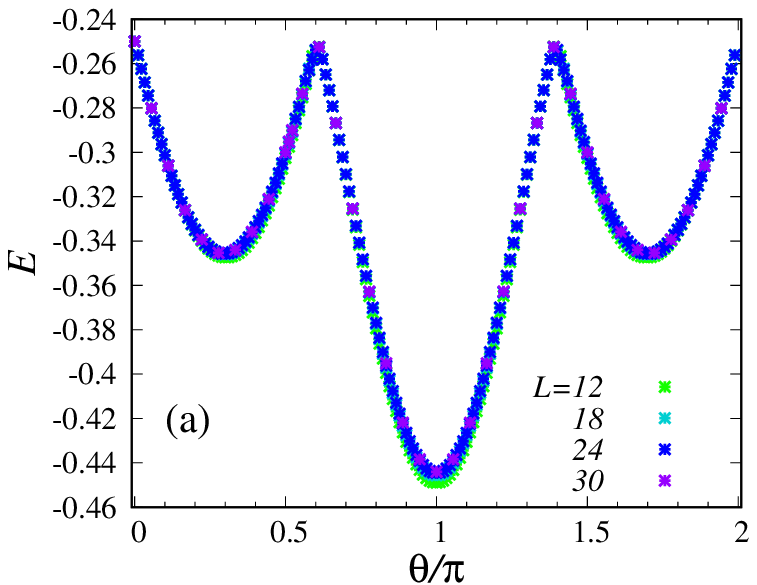}
\includegraphics[width=0.9\hsize]{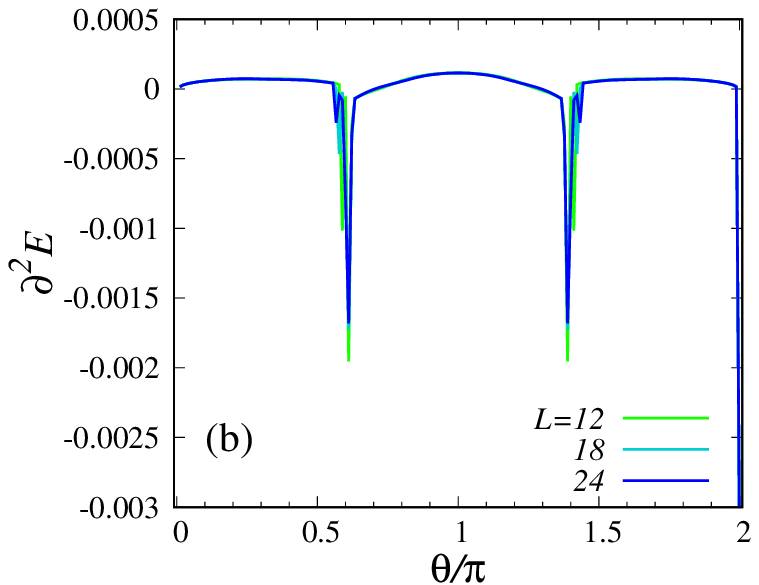}
\vspace{-5mm}
\end{center}
\caption{(Color online) \label{fig:ED_energy} Typical behavior of (a) $E$ and (b) $\partial^2 E/\partial \theta^2$ obtained by the ED.}
\end{figure}
\begin{figure}[htb]
\begin{center}
\includegraphics[width=0.8\hsize]{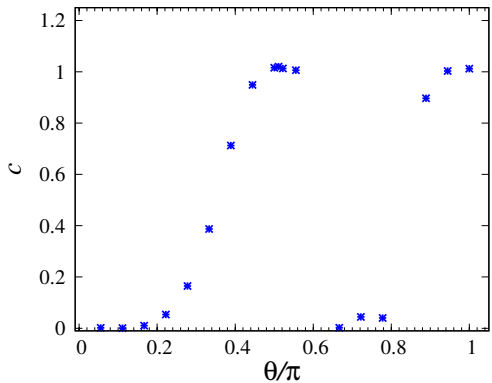}
\end{center}
\vspace{-5mm}
\caption{(Color online) \label{fig:central_charge} Central charge $c$ at $t=1/2$. The central charge $c$ is evaluated from Eq. (5).}
\end{figure}

Figure \ref{fig:ED_energy} illustrates the typical behavior of $E$ and $\partial^2 E/\partial \theta^2$ obtained by the ED as functions of $\theta$.
From these quantities, we recognize three regions distinguished by the first-order phase transition: $0<\theta/\pi<0.611(2)$, $0.611(2)<\theta/\pi< 1.389(2)$, and $1.389(2)<\theta/\pi<2$.
A global $\pi$-rotation around the $S^z$ axis, $({S_i}^x,{S_i}^y,{S_i}^z)\rightarrow(-{S_i}^x,-{S_i}^y,{S_i}^z)$, changes the sign of $\Gamma$, leaving $J$ invariant.
Thus, $E$ satisfies $E(\theta)=E(2\pi-\theta)$, revealing that the phase diagram should be symmetric with respect to $\theta/\pi=1$, and the phases are identical when $0<\theta/\pi < 0.611$ and $1.389 < \theta/\pi<2$.
Consequently, we focus on the results obtained for $0 \le \theta/\pi \le 1$.

In the spin-chain model, we expect a TL liquid at least at the trivial point.
At $\theta/\pi=1$, the system is equivalent to the antiferromagnetic Heisenberg chain; thus, the ground state is the TL liquid.
In Refs. \onlinecite{Affleck1,Affleck2}, it was argued that the TL liquid is stable in the Kitaev-$\Gamma$ spin chain for $\phi_{c1}<\phi \le \phi_{c2}$ and $\phi \ne \pi$, where $K=\cos \phi$, $\Gamma = \sin \phi$, $0<\phi_{c1}/\pi<1/2$, and $3/2<\phi_{c2}/\pi<2$. 
The Kitaev-$\Gamma$ spin chain at $\phi/\pi=3/4$ and $5/4$ is equivalent to the antiferromagnetic Heisenberg spin chain by implementing the six-sublattice rotation \cite{Stavropoulos,Affleck1,Affleck2,Note,QiangLuo}.
When $\phi/\pi \ne 3/4$ or $5/4$, some additional terms that are not renormalized into the Heisenberg term remain after the six-sublattice rotation.
Therefore, it is anticipated that these remaining terms are irrelevant because the TL liquid exists for $\phi_{c1}<\phi \le \phi_{c2}$ and $\phi \ne \pi$.

In the Heisenberg-$\Gamma$ spin chain, the Heisenberg interaction disappears at $\theta/\pi=1/2$ and $3/2$.
Therefore, it can be verified that the ground state at $\theta/\pi=1/2$ and $3/2$ is the TL liquid. 
To elucidate whether or not the TL liquid in the $\Gamma$ (antiferromagnetic Heisenberg) spin chain is also stable against the Heisenberg ($\Gamma$) interaction, 
we evaluate the central charge using the finite-size scaling form based upon the conformal field theory \cite{CFT_FSA1,CFT_FSA2,CFT_FSA3}.
The scaling form is given by
\begin{eqnarray}
\frac{E(L)}{L}=\epsilon_{\infty} -\frac{\pi c v_s}{6L^2},
\label{eq1}
\end{eqnarray}
where $E(L)$ is the ground-state energy for $L$ spins and $c$ is the central charge. 
$\epsilon_{\infty}$ and $v_s$ are the ground-state energy per site and the low-lying excitation velocity in the thermodynamic limit, $L\rightarrow \infty$, respectively.
To evaluate $v_s$, we calculate the lowest energy $E_k(L)$ in the momentum space $k$ and obtain the excitation velocity for $L$ spins from $v_s(L) \approx \frac{L}{2\pi}[E_k(L)-E(L)]$.
By adopting the fitting form, $v_s(L) \approx v_s+\frac{a}{L}+\frac{b}{L^2}$, we estimate $v_s$, where $a$ and $b$ are constants.
Note that we adopt the scaling form Eq. (\ref{eq1}) in $0\le \theta/\pi \le1$ regardless of whether the system shows the linear low-lying dispersion.
The results obtained are shown in Fig. \ref{fig:central_charge}.
We find that the central charge $c=1$ seems to be satisfied for $1/2\le \theta/\pi \lessapprox 0.611$ and in the vicinity of $\theta/\pi=1$.
Thus, the TL liquid characterized by $c=1$ appears in these two regions.
The result in the former region implies that the antiferromagnetic Heisenberg interaction is irrelevant for the TL liquid appearing in the pure $\Gamma$ chain model, similar to the Kitaev interaction in the Kitaev-$\Gamma$ chain model\cite{Affleck1,Affleck2}.
The reason why the antiferromagnetic Heisenberg interaction is irrelevant over such a large portion of the parameter is still to be explained.
Except for the above two regions, $c$ takes nontrivial values. 
These nontrivial values are caused by the fact that the system violates the linear dispersion of the low-lying excitation, which makes it inappropriate to adopt the finite-size scaling form for the ground-state energy given by Eq. (\ref{eq1}). 

\begin{figure}[htb]
\begin{center}
\includegraphics[width=0.9\hsize]{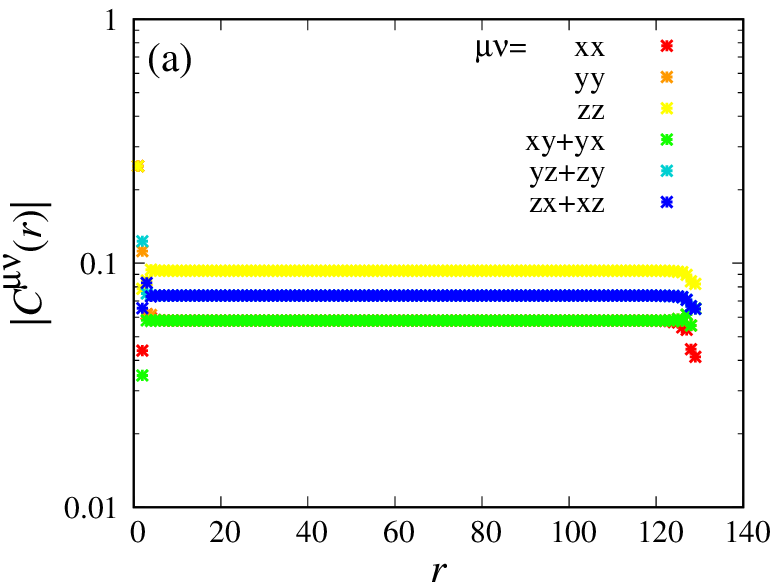}
\includegraphics[width=0.9\hsize]{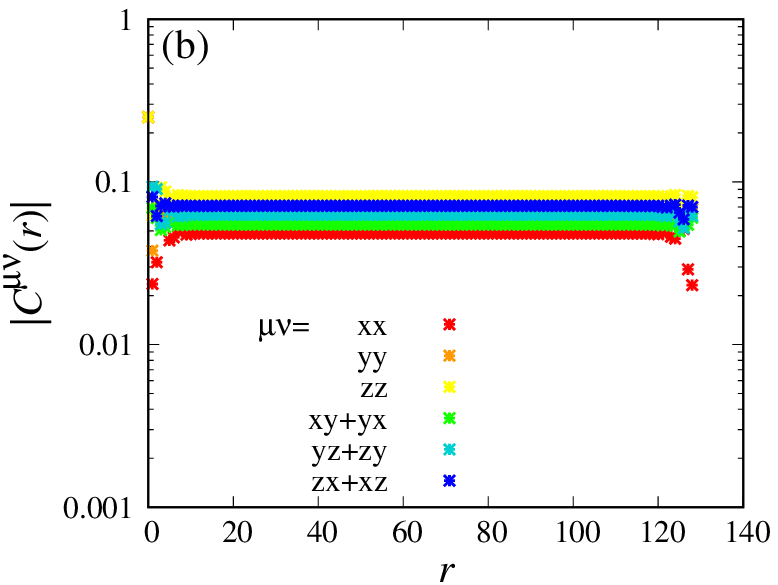}
\vspace{-5mm}
\end{center}
\caption{(Color online) \label{fig:DMRG} Diagonal and off-diagonal spin-spin correlations of the spin-chain model for (a) $\theta/\pi=1/6$ and (b) $\theta/\pi=2/3$. The spin-spin correlation $C^{\mu\nu}(r)$ is obtained by DMRG. We adopt the maximum bond dimension to be $512$.
Note that $C^{\alpha\alpha}=\langle S^{\alpha}(r)S^{\alpha}(0)\rangle$ and $C^{\alpha\beta+\beta\alpha}=\langle S^{\alpha}(r)S^{\beta}(0) +S^{\beta}(r)S^{\alpha}(0) \rangle$/2, where $\alpha,\beta=x, y, z$ and $\alpha \ne \beta $. }
\end{figure}

To elucidate what types of order happens for $0<\theta/\pi<1/2$ and $0.611\le \theta/\pi<1$, we calculate the spin-spin correlation function through DMRG calculations. 
In Fig. \ref{fig:DMRG}, we exhibit the typical results for the $L=256$ chain with the open boundary condition.
We find that for $0<\theta/\pi<1/2$ ($0.611\le \theta/\pi<1$), the spin-spin correlation shows the one-dimensional ferromagnetic (antiferromagnetic) long-range order, because the values of the spin-spin correlation at the longest distance are positively (negatively) finite.
Therefore, we consider that $c=0$ is satisfied for $0\le\theta/\pi<1/2$ and $0.611\le \theta/\pi<1$ in the thermodynamic limit.
Since the phase diagram at $t=1/2$ is symmetric with respect to $\theta/\pi=1$, the TL liquid appears also for $1.389< \theta/\pi\le 3/2$, and the long-range order appears for $1<\theta/\pi \le 1.389$ and $3/2<\theta/\pi<2$.
The emergent TL liquid and two magnetically ordered phases are consistent with the results reported in Ref. \onlinecite{Affleck2}.

\subsection{Isotropically interacting model: $t=1/3$}
\begin{figure}[htb]
\begin{center}
\includegraphics[width=0.9\hsize]{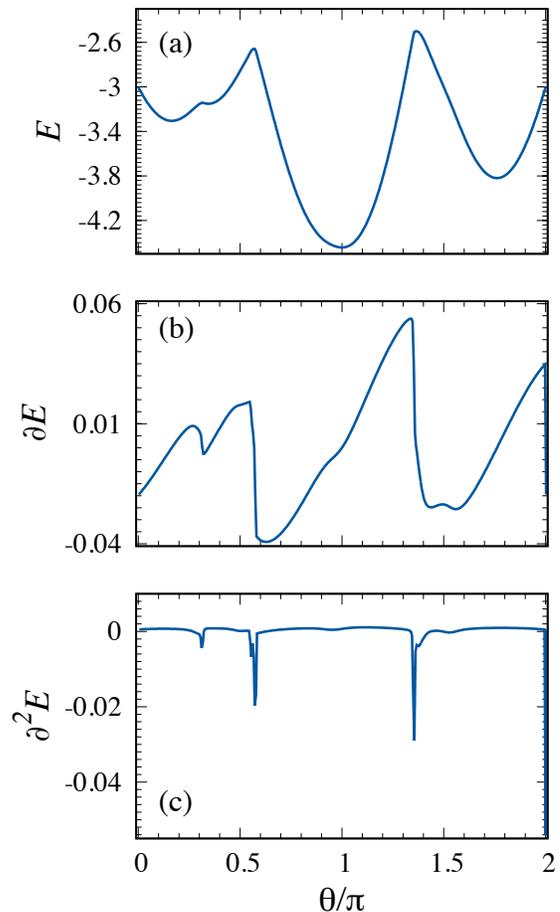}
\vspace{-2mm}
\end{center}
\caption{(Color online) \label{iso_energy} Typical behavior of (a) $E$, (b) $\partial E/\partial \theta$, and (c) $\partial^2 E/\partial \theta^2$ by the 24-site ED at $t=1/3$, where the interactions become spatially isotropic.}
\end{figure}
To investigate the ground state at $t=1/3$, we execute the ED for the 24-site cluster.
In Fig. \ref{iso_energy}, we illustrate the typical behavior of the ground-state energy, $E$, as well as its first and second derivatives, $\partial E/\partial \theta$ and $\partial^2 E/\partial \theta^2$, respectively.
The clear cusps in $E$ reveal that the first-order phase transition takes place at $\theta/\pi=0, 0.317(3), 0.572(3)$, and $1.367(3)$.
Several dips and a broad minimum in $\partial^2 E/\partial \theta^2$ are confirmed at $\theta/\pi \approx 0.556, 1.372$ and $1.528$, respectively.
However, it is a difficult issue to argue whether or not these two dips and this minimum indicate the presence of phase transitions in the thermodynamic limit.
Furthermore calculations and finite-size analysis are needed to clarify this issue.

Further, we focus on the simplest model; at $\theta/\pi=0$ and $1$, the model becomes equivalent to the ferromagnetic and antiferromagnetic Heisenberg models on the honeycomb lattice, respectively.
This indicates that the ferromagnetic (long-range) order appears for $0\le\theta<0.317$ and $1.528<\theta/\pi<2$, and the antiferromagnetic (long-range) order appears for $0.572<\theta/\pi<1.367$.
The static structure factors around $\theta/\pi=0$ and $1$ show prominent peaks at the $\Gamma$ and $\Gamma'$ points in the reciprocal space, respectively (results not shown). 
In the remaining regions, $0.317 < \theta/\pi <0.572$ and $1.372 < \theta/\pi <1.528$, there are no prominent peaks of the static structure factor at symmetric wave vectors, such as the $\Gamma$, $\Gamma'$, $X$, $Y$, and $M$ points.
We consider that the ground state is in the spiral phase which was argued in Ref. \onlinecite{Rau}.

\subsection{$0 < t < 1/3$}
References \onlinecite{Yamaji2018-1,Yamaji2018-2} argued that, by introducing the anisotropy of the interactions as a detour, the Kitaev spin liquid seems to adiabatically connect the $\Gamma$ spin liquid with $|\Gamma/K|\gg1$.
In our previous study\cite{TYamada}, we pointed out that the $\Gamma$ spin liquid in the Kitaev-$\Gamma$ model seems to be unstable against the anisotropy and that the $|t_x\rangle$-dimer state survives up to the isotropically interacting system, $t=1/3$.
Furthermore, in the Kitaev-$\Gamma$ model, the phase boundary at $t=1/3$ is depicted in a large portion of the phase diagram for $\Gamma>0$ and $K<0$ (see Fig. 5(a) in Ref. \onlinecite{TYamada}).
This means that the $|t_x\rangle$-dimer phase is stable when the Kitaev and $\Gamma$ interactions are ferromagnetic and antiferromagnetic, respectively.
In such a case, the long-range magnetically ordered state is suppressed, because the sign combination between the Kitaev and $\Gamma$ interactions induces a strong frustration effect. 
In the Heisenberg-$\Gamma$ model, the frustration effect attributed to the Heisenberg interaction is apparently weaker than that attributed to the Kitaev interaction. 
Thus, whether or not the dimer state in the Heisenberg-$\Gamma$ model is also stable is nontrivial, because the Heisenberg interaction usually favors the magnetic ordering.
To elucidate the stability of the dimer state in the Heisenberg-$\Gamma$ model, we perform dimer series expansions for the Hamiltonian (\ref{Ham2}).

\begin{figure*}[htb]
\begin{center}
\includegraphics[width=0.3\hsize]{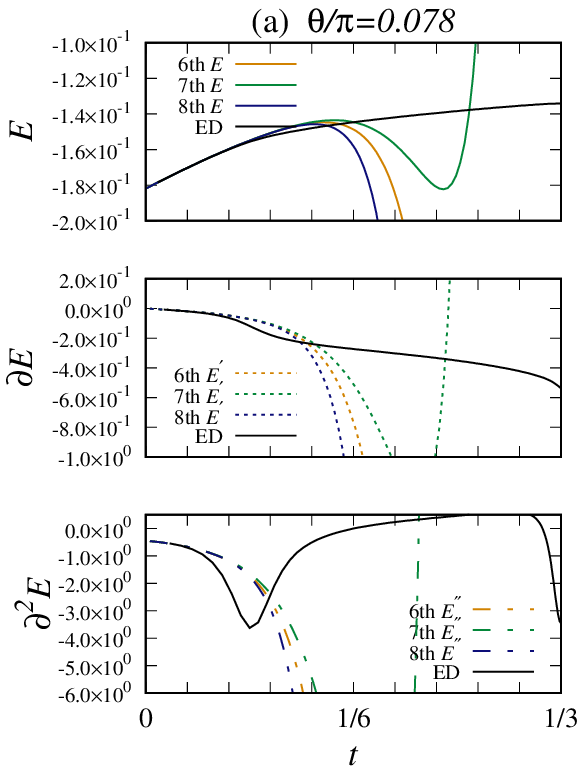}
\includegraphics[width=0.3\hsize]{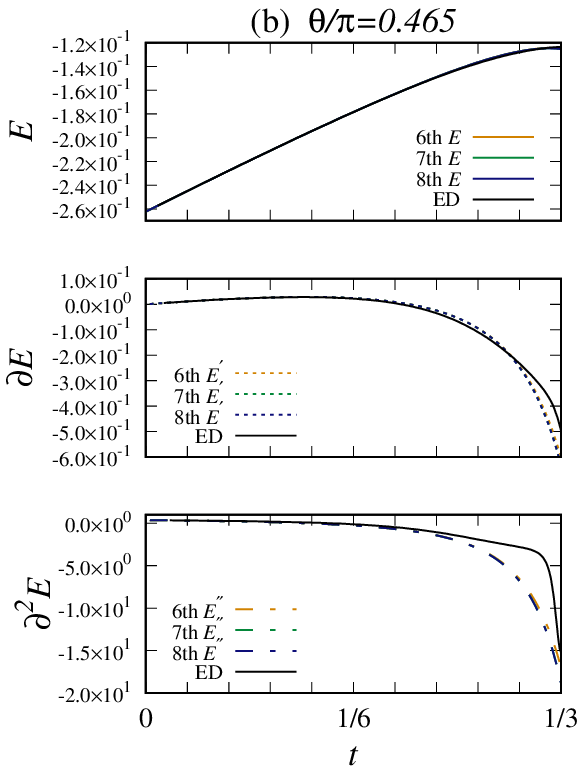}
\vspace{1pt}
\includegraphics[width=0.3\hsize]{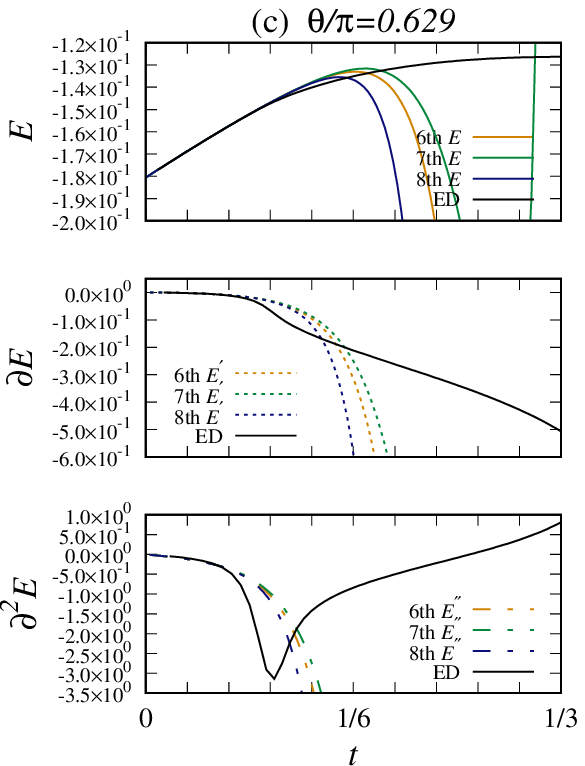}
\includegraphics[width=0.3\hsize]{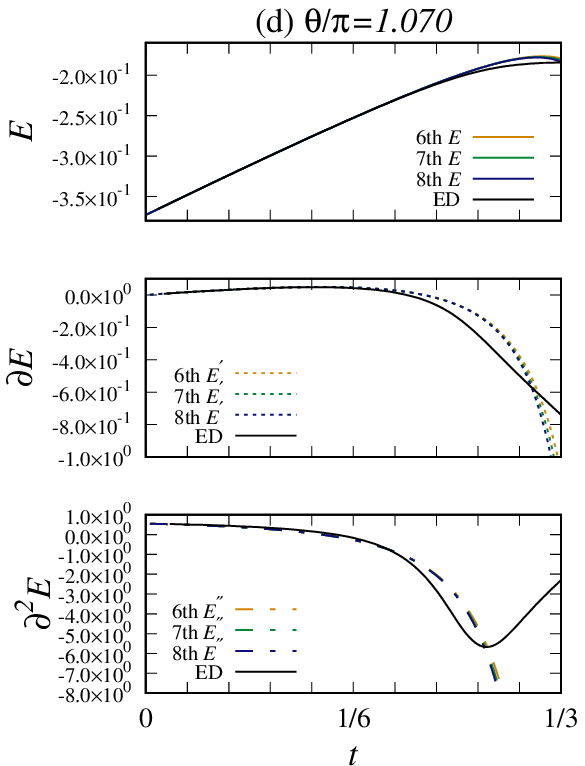}
\vspace{1pt}
\includegraphics[width=0.3\hsize]{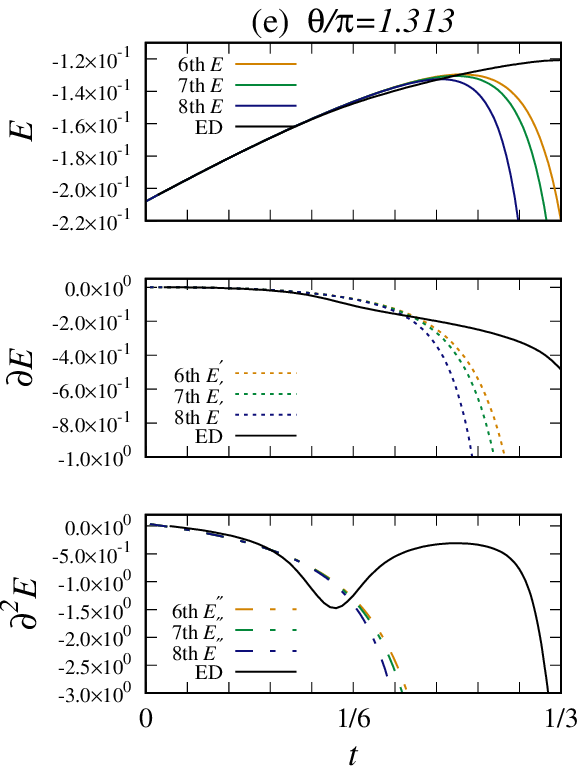}
\includegraphics[width=0.3\hsize]{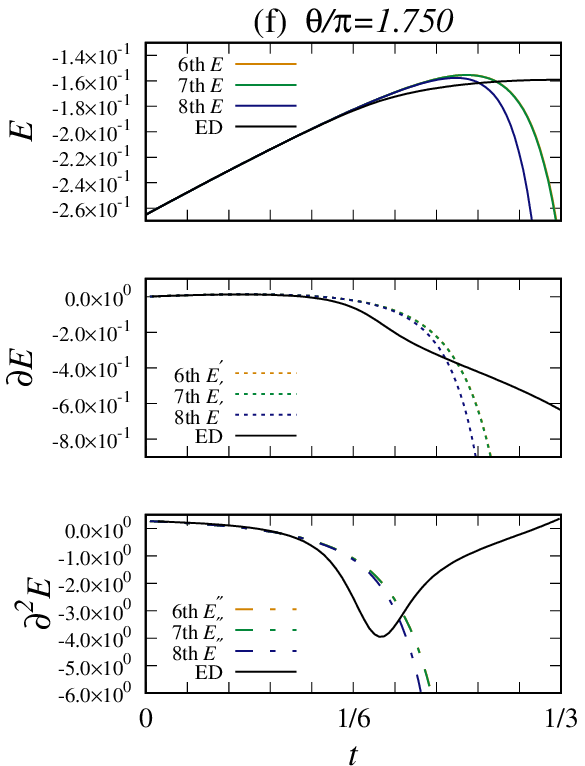}
\hspace{0pc}
\vspace{0pc}
\caption{(Color online) \label{fig:DimerExpansion} Typical behavior of $E$, $\partial E/\partial t$, and $\partial^2 E/\partial t^2$ obtained by dimer series expansions (up to sixth, seventh, and eighth orders) and the 24-site ED. 
  $\theta/\pi={\rm(a)}\; 0.078$, ${\rm(b)}\; 0.465$, ${\rm(c)}\; 0.629$, ${\rm(d)}\; 1.070$, ${\rm(e)}\; 1.313$, and ${\rm(f)}\; 1.750$. }
\end{center}
\end{figure*}

In Fig. \ref{fig:DimerExpansion}, we show the $E$, $\partial E/\partial t$, and $\partial^2 E/\partial t^2$ variation with $t$.
Figures \ref{fig:DimerExpansion}(a)--\ref{fig:DimerExpansion}(f) show the typical behavior of these quantities (up to the sixth--eighth orders) for $\theta/\pi=0.078, 0.465, 0.629, 1.070, 1.313$, and $1.750$.
As the initial state, we adopt the $|t_x\rangle$ dimer [Figs. \ref{fig:DimerExpansion}(a)--\ref{fig:DimerExpansion}(c)], the $|s\rangle$ dimer [Figs. \ref{fig:DimerExpansion}(d) and \ref{fig:DimerExpansion}(e)], and the $|t_y\rangle$ dimer [Fig. \ref{fig:DimerExpansion}(f)].
We also calculate the same quantities by the 24-site ED and compare the results.

Except for Fig. \ref{fig:DimerExpansion}(b), the results obtained by the dimer series expansions show a similar behavior: the ground-state energy convergence becomes worse toward $t=1/3$.
As shown in Figs. \ref{fig:DimerExpansion}(a) and \ref{fig:DimerExpansion}(c)--\ref{fig:DimerExpansion}(f), the ground-state energy exhibits a good convergence for small $t$ and agrees with the results obtained by the 24-site ED.
However, if $t$ is increased, both calculations generate results that start to deviate around $t^*$ where $\partial^2 E/\partial t^2$ obtained by the 24-site ED shows a minimum.
Furthermore, for $t > t^*$, the ground-state energies obtained by the dimer series expansions up to the sixth--eighth orders deviate from each other with increasing $t$, which means that the dimer series expansions fail to properly characterize this model.
We consider that the first-/second-order phase transitions occur at $t \approx t^* (\equiv t_c)$ and the initial dimer state becomes unstable for $t>t^*$. 
Comparing the results of the dimer series expansion with those of the 24-site ED calculations at various $\theta$'s, we find that, at least, both $|s\rangle$-dimer and $|t_y\rangle$-dimer phases disappear at $t \approx  t^*<1/3$ and do not survive up to $t=1/3$. 
These phase transitions in $1/2<\theta/\pi<1$ and $3/2<\theta/\pi<2$ are attributed to a reduction of the frustration owing to the same signs of the Heisenberg and $\Gamma$ interactions.

In contrast, at $\theta/\pi=0.465$ [Fig. \ref{fig:DimerExpansion}(b)], the ground-state energies obtained for up to sixth, seventh, and eighth orders converge for $0\le t \le1/3$, in good agreement with those obtained by the 24-site ED.
The second derivative $\partial^2 E/\partial t^2$ obtained by the 24-site ED shows a steep dip at $t=1/3$, suggesting that, at $t=1/3$, the system is located on the phase boundary. 
We perform the same calculations for various $\theta$'s around $\theta=0.465$ and conclude that the $|t_x\rangle$-dimer phase for $0.32 \lessapprox \theta/\pi \lessapprox 0.57$ survives up to $t=1/3$.

\subsection{$1/3 < t < 1/2$}
\begin{figure*}[htb]
\begin{center}
\includegraphics[width=0.45\hsize]{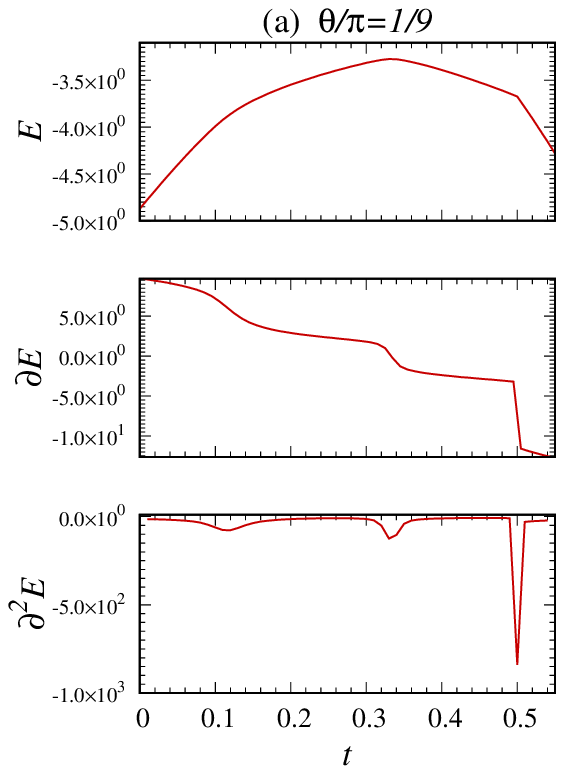}
\includegraphics[width=0.45\hsize]{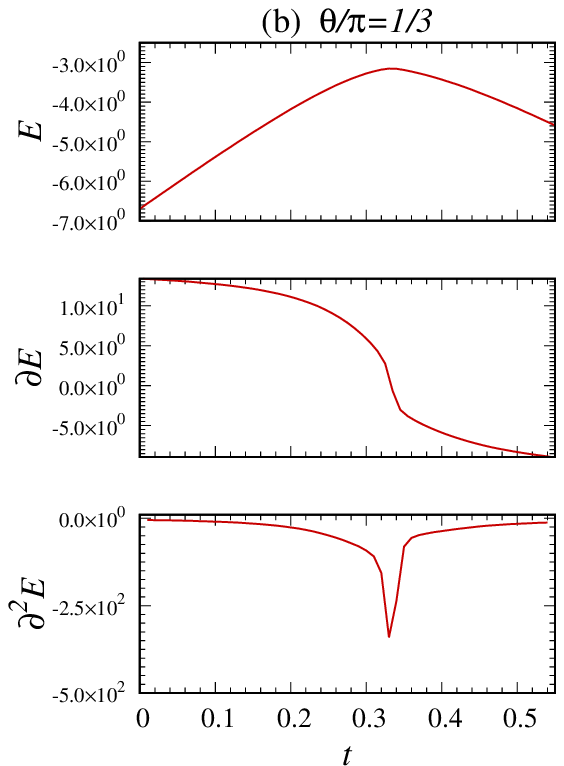}
\hspace{0pc}
\vspace{0pc}
\caption{(Color online) \label{fig:EDChain_to_iso} Typical behavior of $E$, $\partial E/\partial t$, and $\partial^2 E/\partial t^2$ obtained by the 24-site ED. 
  $\theta/\pi={\rm(a)}\; 1/9$ and  ${\rm(b)}\; 1/3$.}
\end{center}
\end{figure*}
For $1/3 < t < 1/2$, we explore the ground-state phase diagram using the Hamiltonian (\ref{Ham1}).
In Fig. \ref{fig:EDChain_to_iso}, we present the typical behavior of $E$, $\partial E/\partial t$, and $\partial^2 E/\partial t^2$ obtained by the 24-site ED.
For $0 \le \theta/\pi \lessapprox 0.28$, $0.5 \lessapprox \theta/\pi \lessapprox 1.39$, and $1.67 \lessapprox \theta/\pi \le 2$, the system is considered to be located on the phase boundary of the first-order phase transition at $t=1/2$, because of the jump in $\partial E/\partial t$ [Fig. \ref{fig:EDChain_to_iso}(a)].
The reason for the first-order phase transition is simplistically explained below.

We consider ferromagnetically/antiferromagnetically ordered spin chains that interact each other via interchain coupling. 
When $t$ increases from below $t=1/2$ to above $t=1/2$, the sign of the interchain coupling changes at $t=1/2$ from positive to negative.
Because of this sign change, the long-range order switches from ferromagnetic/antiferromagnetic order to the conventional $stripe$ pattern.
We expect this phase transition to be of the first-order, because the broken symmetry is different in both phases below and above $t=1/2$.
Thus, we attribute this first-order phase transition to the sign inversion of the interchain spin-spin correlation.

On the other hand, the ground-state energy $E$ for $0.28 \lessapprox \theta/\pi \lessapprox 0.5$ changes continuously, when $t$ decreases from $t=1/2$.
Anomalous behavior, such as a jump or negative divergence, is absent in $\partial E/\partial t$ and $\partial^2 E/\partial t^2$ close to $t=1/2$ [Fig. \ref{fig:EDChain_to_iso}(b)]. 
For $0.28 \lessapprox \theta/\pi \lessapprox 0.5$, the spin-spin correlation along the spin chain composed of the $X$ and $Y$ bonds [see Fig. \ref{Phasediagram}(c)] shows a ferromagnetic long-range order when $t=1/2$; here, the ordering vector is slightly tilted from the diagonal direction of a cube in the spin space, because the amplitudes of the diagonal and off-diagonal spin-spin correlations are different [Fig. \ref{fig:DMRG}(a)].
As $t$ decreases from $1/2$, each ferromagnetically ordered chain starts to correlate via positive $\Gamma$ interactions, which may continuously tilt the direction of the ordering vector in each spin chain.   
The static structure factor $S({\boldsymbol Q})$ shows no prominent peaks for $0.28 \lessapprox \theta/\pi \lessapprox 0.5$ and $1/3<t<1/2$, and the variation of $S({\boldsymbol Q})$ with $t$ is weak [see Fig. \ref{SQ}(c)].
We conclude that the ground state in SP1 corresponds to the spiral phase at $t=1/3$ argued in Ref. \onlinecite{Rau}.
However, the detail of SP1 is still an open question, because as discussed in Refs. \onlinecite{Yamaji2018-1,Yamaji2018-2,Chaloupka_arXiv,GohlkePRR}, the possibility of spin liquid or quantum paramagnet remains.

For $1.35 \lessapprox \theta/\pi \lessapprox 1.65$, the ferromagnetic long-range ordered state and the TL liquid at $t=1/2$ extend slightly toward $t<1/2$, the area of which is noncolored in the phase diagram shown in Fig. \ref{Phasediagram}(d).
We consider that this noncolored area is caused  by the finite-size effect. 
In addition, it is also difficult to clarify the detailed characteristics and phase boundaries of the other noncolored area for $0.2<t<1/3$ around $\theta/\pi=1.37$ in Fig. \ref{Phasediagram}(d).
Further calculations for larger system size are desired to obtain conclusive results.

\section{\label{sec:level5} Summary}
We have investigated the ground-state phase diagram of the Heisenberg-$\Gamma$ model on a honeycomb lattice by using dimer series expansions, ED, and DMRG. 
We have studied the ground-state properties by increasing the interchain interactions, starting from the spin chain limit.  
In the spin chain model except for $0.28<\theta/\pi<0.5$ and $1.39<\theta/\pi<1.67$, the system is located on the phase boundary; moreover, for $1/3<t<1/2$, the two-dimensional magnetic orders are stabilized.
When the interchain interaction is introduced, the two-dimensional magnetic orders, such as F(z), AF, Stripy(z), and F are stabilized and the one-dimensional ferromagnetic long-range order changes to SP1 without a sign of the phase transition.

We have also studied the ground-state properties, starting from the isolated dimer model.
Placing the initial dimers on $Z$ bonds, we have increased the strength of the interactions between nearest-neighbor dimers and investigated the stability of the dimer state. 
We have found that, for a ferromagnetic Heisenberg and antiferromagnetic $\Gamma$ interactions, the $|t_{x}\rangle$-dimer phase is robust and survives up to the isotropically interacting system $t=1/3$ where the phase transition occurs.
This result is similar to what we previously reported \cite{TYamada} for the Kitaev-$\Gamma$ model with ferromagnetic Kitaev and antiferromagnetic $\Gamma$ interactions, where the $|t_{x}\rangle$-dimer phase survived up to the isotropically interacting system. 
Thus, we assert that, for a strong antiferromagnetic-$\Gamma$ interaction, the system favors dimerization.
Furthermore, we expect the $|t_{x}\rangle$-dimer phase to occupy a large part of the phase diagram in the generalized Kitaev model that includes the Heisenberg, $\Gamma$, and ${\Gamma}'$ interactions\cite{Chaloupka_arXiv,Katukuri2014,Winter2016,HSKim2016,WWangPRB}.
In contrast to the $|t_{x}\rangle$-dimer phase, the $|t_y\rangle$-dimer and $|s\rangle$-dimer phases exhibit a phase transition to magnetically ordered  phases at $t_c<1/3$.
This is because the same signs of the Heisenberg and $\Gamma$ interactions causes a reduction of the frustration.

We have summarized the obtained results in the phase diagram.
The obtained phase diagram has indicated that the isotropically interacting system is placed on the phase boundary in a large parameter region, when the system possesses the antiferromagnetic $\Gamma$ interaction.
This implies that the ground state is sensitive to the spatial anisotropy of the interactions. 
Thus, we consider that our results may help to discuss the stability of the ordered state of the isotropically interacting generalized Kitaev model in magnetic fields \cite{HYLeeNature,GohlkePRR,LEChernPRR}.
A semiclassical analysis argued that in the Kitaev-$\Gamma$ model in the magnetic field, the unit cell of the lowest-energy states becomes large when $\Gamma$ increases for $K<0$, and the $C_3$ symmetric state with a fairly large unit cell appears for $0.25<\Gamma/|K|<0.5$\cite{LEChernPRR}.
Such a state is sensitive to the spatial anisotropy of the interactions, because the spatial anisotropy spontaneously breaks the $C_3$ symmetry.
In addition, DMRG results\cite{GohlkePRR} indicated that the cluster geometry used in the computations strongly affects the ordered state.
When the magnetic fields exist, two nematic paramagnet states that breaks $C_3$ rotational symmetry of the lattice were confirmed\cite{GohlkePRR}. 
These two states are competing, which implies that the calculations keeping $C_3$ symmetry of the lattice is important to consider the lowest energy state in the magnetic fields.

\begin{acknowledgments}
This work was supported by the CDMSI, CBSM2, Kinoshita's Foundation, and KAKENHI (Grants No. 19K03721 and No. 21K03390) from MEXT, Japan. 
We are also grateful for the numerical resources at the ISSP Supercomputer Center at the University of Tokyo  
and the Research Center s Nano-Micro Structure Science and Engineering at University of Hyogo.
\end{acknowledgments}


\begin{thebibliography}{99}
\bibitem{Kitaev} 
A. Kitaev, Ann. Phys. (N.Y.) {\bf 321}, 2 (2006). 
%
%
%
\bibitem{Khaliullin}
J. Chaloupka, G. Jackeli and G. Khaliullin, Phys. Rev. Lett.  {\bf 105}, 027204 (2010).
%
\bibitem{SKChoi}
S. K. Choi, R. Coldea, A. N. Kolmogorov, T. Lancaster, I. I. Mazin, S. J. Blundell, P. G. Radaelli, Y. Singh, P. Gegenwart, K. R. Choi, S.-W. Cheong, P. J. Baker, C. Stock, and J. Taylor, Phys. Rev. Lett. {\bf 108}, 127204 (2012).
\bibitem{FYe}
F. Ye, S. Chi, H. Cao, B. C. Chakoumakos, J. A. Fernandez-Baca, R. Custelcean, T. F. Qi, O. B. Korneta, and G. Cao, Phys. Rev. B {\bf 85}, 180403(R) (2012).
\bibitem{JChaloupka}
J. Chaloupka, G. Jackeli, and G. Khaliullin, Phys. Rev. Lett. {\bf 110}, 097204 (2013).
\bibitem{RComin}
R. Comin, G. Levy, B. Ludbrook, Z.-H. Zhu, C. N. Veenstra, J. A. Rosen, Y. Singh, P. Gegenwart, D. Stricker, J. N. Hancock,D. van der Marel, I. S. Elfimov, and A. Damascelli, Phys. Rev. Lett. {\bf 109}, 266406 (2012).
\bibitem{Takagi}
H. Takagi, T. Takayama, G. Jackelli, G. Khaliullin, and S. E. Nagler, Nat. Rev. Phys. {\bf 1}, 264 (2019).
\bibitem{Plump} 
K. W. Plumb, J. P. Clancy, L. J. Sandilands, V. V. Shankar, Y. F. Hu,  K. S. Burch, H.-Y. Kee, and Y.-J. Kim, Phys. Rev. B {\bf 90}, 041112(R) (2014). 
\bibitem{Kubota} 
Y. Kubota, H. Tanaka, T. Ono, Y. Narumi, and K. Kindo, Phys. Rev. B {\bf 91}, 094422 (2015).
\bibitem{Sandilands}
L. J. Sandilands, Y. Tian, K. W. Plumb, Y.-J. Kim, and K. S. Burch, Phys. Rev. Lett. {\bf 114}, 147201 (2015). 
\bibitem{Do}
S.-H. Do, S.-Y. Park, J. Yoshitake, J. Nasu, Y. Motome, Y.-S. Kwon, D. T. Adroja, D.-J. Voneshen, K. Kim, T.-H. Jang, J.-H. Park, K.-Y. Choi, and S. Ji, Nat. Phys. {\bf 13}, 1079 (2017).
\bibitem{DHirobe}
D. Hirobe, M. Sato, Y. Shiomi, H. Tanaka, E. Saitoh, Phys. Rev. B {\bf 95}, 241112(R) (2017). 
\bibitem{Sears} 
J. A. Sears, M. Songvilay, K. W. Plumb, J. P. Clancy, Y. Qiu, Y. Zhao, D. Parshall, and Y.-J. Kim, Phys. Rev. B {\bf 91}, 144420 (2015).
\bibitem{Banerjee1}
A. Banerjee, C. A. Bridges, J.-Q. Yan, A. A. Aczel, L. Li, M. B. Stone, G. E. Granroth, M. D. Lumsden, Y. Yiu, J. Knolle, S. Bhattacharjee, D. L. Kovrizhin, R. Moessner, D. A. Tennant, D. G. Mandrus, and S. E. Nagler, Nat. Mater. {\bf 15}, 733 (2016).
\bibitem{Banerjee2} 
A. Banerjee, J. Yan, J. Knolle, C. A. Bridges, M. B. Stone, M. D. Lumsden, D. G. Mandrus, D. A. Tennant,  R. Moessner, and S. E. Nagler,  Science {\bf 356}, 1055 (2017).
\bibitem{Banerjee3}
A. Banerjee, P. Lampen-Kelly, J. Knolle, C. Balz, A. A. Aczel, B. Winn, Y. Liu, D. Pajerowski, J.-Q. Yan, C. A. Bridges, A. T. Savici, B. C. Chakoumakos, M. D. Lumsden, D. A. Tennant, R. Moessner, D. G. Mandrus, S. E. Nagler, npj Quantum Materials {\bf 3}, 8 (2018).
%
\bibitem{Katukuri2014}
V. M. Katukuri, S. Nishimoto, V. Yushankhai, A. Stoyanova,H.  Kandpal,  S.  Choi,  R.  Coldea,  I.  Rousochatzakis,L. Hozoi, and J. van den Brink, New. J. Phys. {\bf 16}, 013056 (2014).
\bibitem{Yamaji2014}
Y. Yamaji, Y. Nomura, M. Kurita, R. Arita, and M. Imada, Phys. Rev. Lett. {\bf 113}, 107201 (2014).
\bibitem{Sizyuku2014}
Y. Sizyuk, C. Price, P. W\"olfle, and N. B. Perkins, Phys. Rev. B {\bf 90}, 155126 (2014).
\bibitem{Winter2016} 
S. M. Winter, Y. Li, H. O. Jeschke, and R. Valenti, Phys. Rev. B {\bf 93}, 214431 (2016).
\bibitem{WinterRev}
S. M. Winter, A. A. Tsirlin , M. Daghofer, J. van den Brink, Y. Singh, P. Gegenwart, and R. Valent\'{i}, J. Phys.: Condens. Matter {\bf 29}, 493002 (2017).
\bibitem{LJanssenPRB}
L. Janssen, E. C. Andrade, and M. Vojta, Phys. Rev. B {\bf 96}, 064430 (2017).
\bibitem{Majumder2015}
M. Majumder, M. Schmidt, H. Rosner, A. A. Tsirlin, H. Yasuoka,and M. Baenitz, Phys. Rev. B {\bf 91}, 180401(R) (2015). 
\bibitem{Sandilands2016}
 L. J. Sandilands, Y. Tian, A. A. Reijnders, H.-S. Kim, K. W. Plumb, Y.-J. Kim, H.-Y. Kee, and K. S. Burch, Phys. Rev. B {\bf 93}, 075144 (2016).
\bibitem{HSKim2016}
H.-S. Kim, and H.-Y. Kee, Phys. Rev. B {\bf 93}, 155143 (2016).
\bibitem{WWangPRB}
W. Wang, Z.-Y. Dong, S.-L. Yu, and J.-X. Li, Phys. Rev. B {\bf 96}, 115103 (2017).
\bibitem{SuzukiPRB2018}
T. Suzuki and S.-I. Suga, Phys. Rev. B {\bf 97}, 134424 (2018).
\bibitem{PLaurell2020}
P. Laurell, and S. Okamoto, npj Quantum Mater. {\bf 5}, 2 (2020).
\bibitem{Yamaji2018-1} 
A. Catuneanu, Y. Yamaji, G. Wachtel, Y.-B. Kim, and H.-Y. Kee, npj Quantum Materials {\bf 3}, 23 (2018).
\bibitem{Yamaji2018-2} 
M. Gohlke, G. Wachtel, Y. Yamaji, F. Pollmann, and Y. B. Kim, Phys. Rev. B {\bf 97}, 075126 (2018).
\bibitem{WangPRL2019}
J. Wang, B. Normand, and Z.-X. Liu, Phys. Rev. Lett. {\bf 123}, 197201 (2019). 
\bibitem{TYamada}
T. Yamada, T. Suzuki, and S.-I. Suga, Phys. Rev. B {\bf 102}, 024415 (2020). 
\bibitem{textbook}
J. Oitmaa, C. Hamer, and W. Zheng, Series Expansion Methods for Strongly Interacting Lattice Models, Cambridge University Press, New York, 2006.  
\bibitem{JOitmaa}
J. Oitmaa, Phys. Rev B  {\bf 92}, 020405(R) (2015).
\bibitem{Affleck1}  
W. Yang, A. Nocera, T. Tummuru, H.-Y. Kee, and I. Affleck, Phys. Rev. Lett. {\bf 124}, 147205 (2020).
\bibitem{Affleck2} 
W. Yang, A. Nocera, and I. Affleck, Phys. Rev. Research {\bf 2}, 033268 (2020).
\bibitem{Stavropoulos}
P. P. Stavropoulos, A. Catuneanu, and H.-Y. Kee, Phys. Rev. B {\bf 98}, 104401 (2018).
\bibitem{Note}
For $\Gamma>0$ and $K<0$, the system can be mapped onto $(-\Gamma,K)$ by the three sub-lattice rotation argued in Ref. \onlinecite{Affleck1}. Thus, the system satisfying $\Gamma=K<0$ or $-\Gamma=K<0$ is equivalent to the antiferromagnetic Heisenberg spin chain.
\bibitem{QiangLuo}
Q. Luo, J. Zhao, X. Wang, and H.-Y. Kee, Phys. Rev. B {\bf 103}, 144423 (2021).
\bibitem{CFT_FSA1}
J. L. Cardy, J. Phys. A {\bf 17}, 385 (1984).
\bibitem{CFT_FSA2}
H. W. J. Bl\"ote, J. L. Cardy, and M. P. Naightingale, Phys. Rev. Lett. {\bf 56}, 742 (1986).
\bibitem{CFT_FSA3}
I. Affleck, Phys. Rev. Lett. {\bf 56}, 746 (1986).
\bibitem{Rau}
J. G. Rau, E. K.-H. Lee, and H.-Y. Kee, Phys. Rev. Lett.  {\bf 112}, 077204 (2014). 
\bibitem{Chaloupka_arXiv}
J. Chaloupka and G. Khaliullin, Phys. Rev. B {\bf 92}, 024413 (2015).
\bibitem{GohlkePRR}
M. Gohlke, L. E. Chern, H.-Y. Kee, and Y. B. Kim, Phys. Rev. Research {\bf 2}, 043023 (2020).
\bibitem{HYLeeNature}
H.-Y. Lee, R. Kaneko, L. E. Chern, T. Okubo, Y. Yamaji, N. Kawashima, and Y. B. Kim, Nat. Comm. {\bf 11},  1639 (2020).
%
\bibitem{LEChernPRR}
L. E. Chern, R. Kaneko, H.-Y. Lee, and Y. B. Kim, Phys. Rev. Research {\bf 2}, 013014 (2020).
%
%
\end{thebibliography}
\end{document}